\documentclass[traditabstract,longauth]{aa}

\usepackage{graphicx}
\usepackage{txfonts}
\usepackage{natbib}
\usepackage[colorlinks=true,citecolor=blue]{hyperref}
\usepackage{xspace}
\usepackage{subfig}
\usepackage{xcolor}

\usepackage[T1]{fontenc}

\bibpunct{(}{)}{;}{a}{}{,}

\newcommand{\MJup}{\ensuremath{M_{\mathrm{Jup}}}\xspace}

\newcommand{\mic}{$\mu$m\xspace}
\newcommand{\as}{\hbox{$^{\prime\prime}$}\xspace}

\begin{document}

\title{First light of the VLT planet finder SPHERE}
\subtitle{III. New spectrophotometry and astrometry of the HR\,8799 exoplanetary system \thanks{Based on observations collected at the European Southern Observatory (ESO), Chile, during the commissioning of the SPHERE instrument and during the science verification (program number 60.A-9352(A)).}}

   \author{A. Zurlo\inst{1,2,3,4}, A. Vigan\inst{3,5}, R. Galicher\inst{6},  A.-L. Maire\inst{4}, D. Mesa\inst{4}, R. Gratton\inst{4}, G. Chauvin\inst{7,8}, M. Kasper\inst{9,7,8}, C. Moutou\inst{3}, M. Bonnefoy\inst{7,8}, S. Desidera\inst{4}, L. Abe\inst{10}, D. Apai\inst{11,12,13}, A. Baruffolo\inst{4}, P. Baudoz\inst{6}, J. Baudrand\inst{6}, J.-L. Beuzit\inst{7,8}, P. Blancard\inst{3}, A. Boccaletti\inst{6}, F. Cantalloube\inst{7,8}, M. Carle\inst{3}, E. Cascone\inst{14}, J. Charton\inst{8}, R.U. Claudi\inst{4}, A. Costille\inst{3}, V. de Caprio\inst{14}, K. Dohlen\inst{3}, C. Dominik\inst{15}, D. Fantinel\inst{4}, P. Feautrier\inst{8}, M. Feldt\inst{16}, T. Fusco\inst{3,17}, P. Gigan\inst{6}, J.H. Girard\inst{5,7,8}, D. Gisler\inst{18}, L. Gluck\inst{7,8}, C. Gry\inst{3}, T. Henning\inst{16}, E. Hugot\inst{3}, M. Janson\inst{19,}\inst{16}, M. Jaquet\inst{3}, A.-M. Lagrange\inst{7,8}, M. Langlois\inst{20,3}, M. Llored\inst{3}, F. Madec\inst{3}, Y. Magnard\inst{8}, P. Martinez\inst{10}, D. Maurel\inst{8}, D. Mawet\inst{21}, M.R. Meyer\inst{22}, J. Milli\inst{5,7,8}, O. Moeller-Nilsson\inst{16}, D. Mouillet\inst{7,8}, A. Orign\'e\inst{3}, A. Pavlov\inst{16}, C. Petit\inst{17}, P. Puget\inst{8}, S.P. Quanz\inst{22}, P. Rabou\inst{8}, J. Ramos\inst{16}, G. Rousset\inst{6}, A. Roux\inst{8}, B. Salasnich\inst{4}, G. Salter\inst{3}, J.-F. Sauvage\inst{3,17}, H.M. Schmid\inst{22}, C. Soenke\inst{5}, E. Stadler\inst{8}, M. Suarez\inst{5}, M. Turatto\inst{4}, S. Udry\inst{23}, F. Vakili\inst{10}, Z. Wahhaj\inst{5}, F. Wildi\inst{23}, J. Antichi\inst{24}}

 \institute{\inst{1} N\'ucleo de Astronom\'ia, Facultad de Ingenier\'ia, Universidad Diego Portales, Av. Ejercito 441, Santiago, Chile \\
   \inst{2} Departamento de Astronom\'ia, Universidad de Chile, Casilla 36-D, Santiago, Chile \\
   \inst{3} Aix-Marseille Universit\'e, CNRS, LAM (Laboratoire d'Astrophysique de Marseille) UMR 7326, 13388, Marseille, France\\
\inst{4} INAF - Osservatorio Astronomico di Padova, Vicolo dell'Osservatorio 5, 35122, Padova, Italy \\
     \inst{5} European Southern Observatory, Alonso de Cordova 3107, Vitacura, Santiago, Chile \\ 
   \inst{6} LESIA, Observatoire de Paris, CNRS, Universit\'e Pierre et Marie Curie - Paris 6 \& Universit\'e Paris Diderot - Paris 7, 5 place Jules Janssen, F-92190 Meudon, France \\
\inst{7} Universit\'e Grenoble Alpes, IPAG, 38000 Grenoble, France \\
\inst{8} CNRS, IPAG, F-38000 Grenoble, France\\
\inst{9} European Southern Observatory, Karl-Schwarzschild-Strasse 2, D-85748 Garching, Germany\\
\inst{10} Laboratoire Lagrange, UMR7293, Universit\'e de Nice Sophia-Antipolis, CNRS, Observatoire de la Cote d'Azur, Bd. de l'Observatoire, 06304 Nice, France \\
\inst{11} Steward Observatory, Department of Astronomy, University of Arizona, 993 North Cherry Avenue, Tucson, AZ 85721, USA\\
\inst{12} Lunar and Planetary Laboratory, University of Arizona, 1640 E. Univ. Blvd., Tucson, USA \\
\inst{13} Earth in Other Solar Systems Team \\
\inst{14} INAF - Astrophysical Observatory of Capodimonte, Salita Moiariello 16, 80131 Napoli, Italy \\
\inst{15} Anton Pannekoek Institute for Astronomy, University of Amsterdam, PO Box 94249, 1090 GE Amsterdam, The Netherlands \\
\inst{16} Max-Planck-Institut f\"ur Astronomie, K\"onigstuhl 17, 69117 Heidelberg, Germany\\
\inst{17} Office National d'Etudes et de Recherches A\'erospatiales, 29 avenue de la division Leclerc, 92322 Ch\^atillon  \\
\inst{18} Kiepenheuer-Institut für Sonnenphysik, Schöneckstr. 6, D-79104 Freiburg, Germany \\
\inst{19} Stockholm University, AlbaNova University Center, Stockholm, Sweden \\
\inst{20} CRAL, UMR 5574, CNRS, Universit\'e Lyon 1, 9 avenue Charles Andr\'e, 69561 Saint Genis Laval Cedex, France\\
\inst{21} Department of Astronomy, California Institute of Technology, 1200 E. California Blvd, MC 249-17, Pasadena, CA 91125 USA \\
\inst{22} Institute for Astronomy, ETH Zurich, Wolfgang-Pauli-Strasse 27, 8093 Zurich, Switzerland \\
\inst{23} Observatoire de Gen\`eve, University of Geneva, 51 Chemin des Maillettes, 1290, Versoix, Switzerland \\
 \inst{24} INAF - Osservatorio Astrofisico di Arcetri, Largo E. Fermi 5, 50125 Firenze, Italy }           
   \date{Submitted/Accepted}


  \abstract
 {The planetary system discovered around the young A-type HR\,8799 provides a unique laboratory to: a) test planet formation theories, b) probe the diversity of system architectures at these separations, and c) perform  comparative (exo)planetology.}
    {We present and exploit new near-infrared images and integral-field spectra of the four gas giants surrounding HR\,8799 obtained with SPHERE, the new planet finder instrument at the Very Large Telescope, during the commissioning and science verification phase of the instrument (July-December 2014). With these new data, we contribute to completing the spectral energy distribution of these bodies in the 1.0-2.5 $\mu$m range. We also provide new astrometric data, in particular for planet e, to further constrain the orbits.}
     { We used the infrared dual-band imager and spectrograph (IRDIS) subsystem to obtain pupil-stabilized, dual-band $H2H3$ (1.593 $\mu$m, 1.667 $\mu$m), $K1K2$ (2.110 $\mu$m, 2.251 $\mu$m), and broadband $J$ (1.245 $\mu$m) images of the four planets. IRDIS was operated in parallel with the integral field spectrograph (IFS) of SPHERE to collect low-resolution ($R\sim30$), near-infrared (0.94-1.64 $\mu$m) spectra of the two innermost planets HR\,8799\,d and e \thanks{The spectra are only available in electronic form at the CDS via \url{http://cdsweb.u-strasbg.fr/cgi-bin/qcat?J/A+A/}}.
      The data were reduced with dedicated algorithms, such as the Karhunen-Lo\`eve image projection (KLIP), to reveal the planets. We used the so-called negative planets injection technique to extract their photometry, spectra, and measure their positions. We illustrate the astrometric
performance of SPHERE through sample orbital fits compatible with 
SPHERE and literature data.}
   {We demonstrated the ability of SPHERE to detect and characterize planets in this kind of systems, providing spectra and photometry of its components. The spectra improve upon the signal-to-noise ratio of previously obtained data and increase the spectral coverage down to the $Y$ band. In addition, we provide the first detection of planet e in the $J$ band. Astrometric positions for planets HR\,8799\,bcde are reported for the epochs of July, August, and December 2014.  We measured the photometric values in $J$, $H2H3$, $K1K2$ bands for the four planets with a mean accuracy of 0.13 mag. We found upper limit constraints on the mass of a possible planet f of 3-7~\MJup.  Our new measurements are more consistent with the two inner planets d and e being in a 2d:1e or 3d:2e resonance.  The spectra of HR\,8799\,d and e are well matched by those of L6-8 field dwarfs. However, the SEDs of these objects are redder than field L dwarfs longward of 1.6 $\mu$m.}
   {}


\titlerunning{SPHERE results on the HR\,8799 planetary system.}
\authorrunning{Zurlo et al.}
\maketitle
%
\section{Introduction}

The intriguing planetary system around HR\,8799 is one of the most interesting objects in the field of extrasolar planets. Four giant planets have been detected \citep[HR\,8799\,bcde;][]{2008Sci...322.1348M,2010Natur.468.1080M} as well as a double debris belt \citep{2009ApJ...705..314S,2011ApJ...740...38H,2014ApJ...780...97M}.

HR\,8799 is a $\gamma$ Dor-type variable star \citep{1999AJ....118.2993G} with $\lambda$ Boo-like abundance patterns.   Its age has been estimated to be 20-160~Myr \citep{1969AJ.....74..375C, 2006ApJ...644..525M,2008Sci...322.1348M, 2010ApJ...716..417H,2011ApJ...732...61Z,2012ApJ...761...57B}, and the distance of the system is $39.4 \pm 1.0$~pc \citep{2007A&A...474..653V}. This system is a benchmark for the study of young, gaseous giant planets, their formation and evolution. The planets have relative low contrast ($\Delta H$ $\sim$ 12 mag) and are separated from their host star (15-70 AU) such that  they are relatively easy to detect with the new generation of high-contrast imagers utilizing optimized image processing techniques. For an assumed age of 30~Myr (Columba association), the measured luminosities of the planets suggest masses around 5-7 \MJup \citep{2010Natur.468.1080M,2011ApJ...729..128C,2012ApJ...755...38S}, in agreement with the upper limits set by dynamical stability studies of the system \citep{2014MNRAS.440.3140G}.

So far, studies of the atmospheres of HR\,8799\,bcde were conducted using mostly photometric measurements in the near infrared. The system has been detected by several instruments, providing broadband photometry and/or astrometry in $z$, $J$, $H$, $K$, $L$, $M$ filters \citep{2008Sci...322.1348M,2009ApJ...694L.148L,2010Natur.468.1080M,2010ApJ...716..417H,2010ApJ...710L..35J,2010AAS...21537706S,2011ApJ...729..128C,2011ApJ...730L..21H,galicher11,2011ApJ...741...55S,2012ApJ...753...14S,2013A&A...549A..52E,2014ApJ...792...17S}.  A new generation of high-contrast imaging instruments such as SPHERE, GPI, ScEXAO+CHARIS, Project 1640 \citep{2008SPIE.7014E..18B, 2014PNAS..11112661M,2012SPIE.8446E..9CM, 2011PASP..123...74H} can now provide the spectra of these objects, enabling deeper study of their chemistry and physical properties.  Spectra of the companions have been obtained in the $JH$ band for planets b, c, d, and e by Project 1640 at Palomar \citep{2013ApJ...768...24O}, in the $H$ and $K$ bands using Keck/OSIRIS \citep{2006SPIE.6269E..1AL} for planets b \citep{2011ApJ...733...65B,2010ApJ...723..850B,2015ApJ...804...61B} and c \citep{2013Sci...339.1398K}, exploiting GPI for planets c and d \citep{2014ApJ...794L..15I}, and in 3.88-4.10 $\mu$m with VLT/NACO for planet c \citep{2010ApJ...710L..35J}.

State-of-the-art analysis of the atmospheres presents a complex picture of the objects of this system and {\it ad hoc} models are needed to fit the spectrophotometric data.  The planets around HR\,8799 show nonequilibrium chemistry in the atmospheres and they appear dustier than field objects of the same effective temperature \citep{2011ApJ...729..128C}. The fact that HR\,8799\,bcde are brighter at 3.3~$\mu$m than predicted by the equilibrium chemistry models optimized for old brown dwarfs\ suggests that the upper layers of their atmosphere are not as opaque as expected because of the lack of CH$_{4}$  \citep{2012ApJ...753...14S}.
To reconcile the discrepancies between models and data, a patchy cloud coverage has been suggested by \citet{2008Sci...322.1348M}, \citet{2011ApJ...729..128C}, \citet{2012ApJ...753...14S}, \citet{2012ApJ...754..135M}, and \citet{2012ApJ...756..172M}. \citet{2011ApJ...737...34M} proposed a set of models that are intermediate between two sets of models; one  where the physical extent of the clouds is truncated at a given altitude, and a second  model where clouds extend all the way to the top of the atmosphere. The fact that we need thick clouds and nonequilibrium chemistry to fit the observables could be related to the low surface gravities of the planets around HR\,8799 \citep{2011ApJ...737...34M, 2012ApJ...754..135M}. This patchy cloud hypothesis is fully consistent with the high-precision rotational phase mapping studies of field brown dwarfs with spectra similar to HR\,8799\,b, demonstrating heterogeneous clouds (warm thin and cool thick clouds) for several objects \citep{2013ApJ...768..121A,2015ApJ...798..127B}.

In this paper, we present new data obtained with VLT/SPHERE of the objects in the system. We obtain  photometry and astrometry in $J$, $H,$ and $K$ bands for the four planets and extract $YH$ band spectra for planets d and e.  We then provide an updated interpretation of the complex atmospheres of the HR\,8799 planets, taking advantage of the exquisite quality of the data obtained. 

The outline of the paper is as follow: in Sect.~\ref{sec:obs} we present SPHERE near-infrared observations of the system around HR\,8799 during the instrument commissioning and science verification runs in July, August, and December 2014; in Sect.~\ref{sec:red} we describe the reduction methods applied and the results that we obtained. In Sect.~\ref{Sec:astro} we present the astrometric fitting for the four planets of HR\,8799. In Sect.~\ref{sec:specphotanalysis} we present a first comparison with data from objects of comparable temperature, while  a more detailed analysis is found in Bonnefoy et al., { A\&A accepted}.  We give the conclusions in Sect.~\ref{sec:conc}.


\section{Observations}
\label{sec:obs}
We observed HR\,8799 over three different nights during the SPHERE commissioning runs and four different nights during science verification. The SPHERE planet-finder instrument installed at the VLT \citep{2008SPIE.7014E..18B} is a highly specialized instrument dedicated to high-contrast imaging.  It was built by a large consortium of European laboratories. It is equipped with an extreme adaptive optics system, SAXO \citep{Fusco:06, 2014SPIE.9148E..0OP}, with a $41\times41$ actuators wavefront control, pupil stabilization, differential tip-tilt control, and it also employs stress polished toric mirrors for beam transportation \citep{2012A&A...538A.139H}. The SPHERE\ instrument is equipped with several coronagraphic devices for stellar diffraction suppression, including apodized Lyot coronagraphs \citep{2005ApJ...618L.161S} and achromatic four-quadrants phase masks \citep{Boc08}. The instrument has three science subsystems: the infrared dual-band imager and spectrograph \citep[IRDIS;][]{Do08}, an integral field spectrograph  \citep[IFS;][]{Cl08} and the Zimpol rapid-switching imaging polarimeter \citep[ZIMPOL;][]{Th08}.

Two sequences of data were acquired with the integral-field spectrometer IFS \citep{2008SPIE.7014E..3EC} and the dual-band imager IRDIS \citep{2008SPIE.7014E..3LD} in dual-band imaging mode \citep{2010MNRAS.407...71V} working in parallel. The last sequence was taken with IRDIS alone in the broadband imaging mode with the $J$ band filter. Finally, a broadband imaging sequence in $H$ band was taken during the science verification (SV) phase. Data in different spectral ranges were obtained in the following configurations:
\begin{itemize}
\item The first sequence was taken on July, 13th 2014 in IRDIFS mode (IRDIS in dual-band imaging mode using the $H2H3$ filters, and IFS in the $YJ$ mode). The seeing was variable between $0.8-1.0\as$. Sixteen datacubes were acquired, with a total exposure time of 43~min. The total field of view (FoV) rotation during the observation was 18.1$^{\circ}$.
 
\item An IRDIFS\_EXT sequence (IRDIS in dual-band imaging mode with the $K1K2$ filters, and IFS in the $YH$ mode) was taken on August, 12th 2014 with good seeing conditions. Thirty-two datacubes were acquired for each instrument with a total exposure time of nearly two hours. The total field rotation was 33.8$^{\circ}$.
\item An IRDIS broadband $J$ sequence was taken on August, 14th 2014. Thirty-two datacubes were acquired for an integration time of $\sim$ 1.4~hours. The total FoV rotation was 31.2$^{\circ}$.
\item Four IRDIS broadband $H$ sequences were taken, between December, 4th to 8th 2014, during SV. Integration times for each data set were around 30 minutes, and field rotations were around 8$^{\circ}$.
\end{itemize}

IRDIS filters and resolutions are listed in Table~\ref{t:filt}. The observations are summarized in Table~\ref{t:obs}.

\begin{table}
\caption{IRDIS filters used during the observations and their resolutions.} 
\label{t:filt}
\centering
\begin{tabular}{lcc}
\hline
\hline
Filter &  Wavelength (\mic) & Resolution \\
\hline
BB\_$J$  &      1.245  &  5 \\
BB\_$H$ &   1.625       & 6  \\
$H2$  & 1.593 & 30 \\
$H3$  &1.667  &  30 \\
$K1$  &2.110  & 20  \\
$K2$   &2.251   & 20 \\
\hline
\end{tabular}
\end{table}

\begin{table*}
\begin{minipage}{1\textwidth}
\caption{Observations of HR\,8799 during SPHERE commissioning and science verification runs.} 
\label{t:obs}
\renewcommand{\footnoterule}{}  
\centering
\begin{tabular}{ccccccc}
\hline
\hline
           UT date           &  IRDIS Filter   & IFS Band & IRDIS DIT\footnote{Detector Integration Time}$\times$NDIT\footnote{Number of frames per dithering position} & IFS DIT$\times$NDIT & N. of datacubes &FoV Rotation  \\
\hline
2014-07-13      & $H2H3$ & $YJ$ & 4$\times$40 & 8$\times$20 & 16&18.1$^{\circ}$\\
2014-08-12      & $K1K2$ & $YH$ & (30+35)$\times$6 & 60$\times$3+100$\times$2 &16+16& 33.8$^{\circ}$\\
2014-08-14      & BB\_$J$ & - & 16$\times$10 & - &32& 31.2$^{\circ}$\\
2014-12-04                      & BB\_$H$ & - & 8$\times$218 & - & 1& 8.7$^{\circ}$\\
2014-12-05                      & BB\_$H$ & - & 8$\times$218 & - & 1& 8.7$^{\circ}$\\
2014-12-06                      & BB\_$H$ & - & 8$\times$218 & - & 1& 8.5$^{\circ}$\\
2014-12-08                      & BB\_$H$ & - & 8$\times$218 & - & 1& 7.9$^{\circ}$\\

\hline
\end{tabular}
\end{minipage}
\end{table*}

All the sequences were taken with the same configuration of the apodized Lyot coronagraph \citep{2011ExA....30...39C}, which  includes a focal plane mask with a diameter of 185~mas and an inner working angle (IWA) of 0\farcs09.

The commissioning sequences were taken with the following observing strategy:
\begin{itemize}
\item A star centered frame, which is a coronagraphic image of the star with four satellite spots symmetric with respect to the central star, was acquired at the beginning and at the end of each observation by introducing a periodic modulation on the deformable mirror. The purpose of this observation is to accurately determine the star position for frame registering before the application of angular differential image (ADI) processing \citep{2006ApJ...641..556M} and for derivation of the astrometry for the detected companions relative to the star \citep{2006ApJ...647..612M, 2006ApJ...647..620S}. { The satellite spots have high SNR ($\sim$ 50), which permits a good estimation of the center with milliarcsecond precision}.
\item Nonsaturated images of the star shifted from the central axis using a tip-tilt mirror were taken at the beginning and at the end of the observations with a neutral density filter, ND2.0, with a transmission $\sim$1\% of the total bandwidth, to calibrate the flux.
\item Coronagraphic images were taken with a 4$\times$4 dithering pattern for IRDIS. These images were obtained in pupil-stabilized mode to take advantage of the angular ADI technique. 
\item At the end of  sequence six, sky background images were obtained with the same exposure time for the coronagraphic and unsaturated point spread function (PSF) images.
\end{itemize}

The broadband $H$ sequences during SV were acquired with the four satellite spots permanently applied.
All the other calibrations described in Sects.~\ref{sec:irdis_red} and~\ref{sec:ifs_red} were acquired during the next morning as part of the instrument calibration plan.

\section{Data reduction, photometry and astrometry}
\label{sec:red}

\subsection{Astrometric calibration}
\label{sec:cal}

{ During each run of observations we observed a dedicated field in the outer region of the globular cluster 47\,Tuc to derive astrometric calibrations of IRDIS images (distortion, plate scale, and orientation), using the Hubble Space Telescope (HST) data as a reference  \citep[A. Bellini \& J. Anderson, private comm.; see][for the methods used to obtain HST measurements]{2014ApJ...797..115B}. A detailed description of the astrometric calibration is presented in \citet{2016A&A...587A..56M}. 

The 47\,Tuc data were reduced and analyzed using the Data Reduction and Handling software \citep[DRH;][]{Pa08} and IDL routines. To calculate the position of each star in the FoV, we used the IDL function \textit{cntrd}\footnote{http://idlastro.gsfc.nasa.gov/ftp/pro/idlphot/} from the DAOphot software \citep{1987PASP...99..191S}. We derived our IRDIS calibrations from the comparison between the measured positions on the detector of several tenths of stars and the HST observations (March 13, 2006).
The typical accuracy of the catalog positions is $\sim 0.3$~mas and we took into account the time baseline between the HST and SPHERE observations.  The distortion measured on sky is dominated by an anamorphism factor of ($0.60\pm 0.02) \%$ between the horizontal and  vertical directions of the detector. 
The anamorphism is common to both instruments because it is produced by cylindrical mirrors in the common path and infrastructure of the instrument.

Concerning IFS, as we did not obtain observations of any relevant astrometric calibrator, we calculated plate scale and orientation with a simultaneous observation of a distortion grid on both IRDIS and IFS detectors. We estimated for the IFS data a plate scale of $7.46\pm 0.01$ mas/pix and a relative orientation to IRDIS data of $-100.46 \pm 0.13 \deg$ .

The astrometric calibrator sequences { of 47\,Tuc} were acquired during the same observing runs as the science observations reported in this paper and with the same instrument setup (filter, coronagraph). Values calculated for the plate scale and for  true north are shown in Table~\ref{t:plate}.

Raw data of both IRDIS and IFS have been corrected for anamorphism before the post-processing, taking into account that the IFS detector is rotated with respect to the IRDIS one.

\begin{table}
\caption{Values of the plate scale and the true north for IRDIS observations measured during in each run of the observations. Values of July and August 2014 have been measured for the $H2H3$ filter, while December values refer to BB\_$H$ band filter. The measurements for the left and right parts of the IRDIS detector are shown. } 
\label{t:plate}
\centering
\begin{tabular}{lcc}
\hline
\hline
Date & Plate scale (mas/px) & True north (deg)\\
\hline
\multicolumn{3}{c}{\it Left field}\\
\hline
July 2014     & $12.252\pm0.006$  &     $-1.636\pm0.013$  \\
August 2014   &  $12.263\pm0.006$       & $-1.636\pm0.013$  \\
December 2014   & $12.251\pm0.005$      & $-1.764\pm0.010$   \\
\hline
\multicolumn{3}{c}{\it Right field}\\
\hline
July 2014     &$12.247\pm0.006$  &      $-1.653\pm0.005$ \\
August 2014    &  $12.258\pm0.006$      & $-1.653\pm0.005$  \\
December 2014 & $12.241\pm0.004$        & $-1.778\pm0.010$  \\
\hline
\end{tabular}
\end{table}

The pixel scale variations from one observing run to another can be ascribed to ambient temperature variations. If we consider astrometric calibrator observations from the commissioning runs, we found the following relation between the pixel scale and the temperature: $scale~$(mas/px)$ = 0.0060~ \cdot$ T$_{amb}$($^{\circ}$C)$ + 12.1655$. The residual scatter is 0.012 mas/px. This relation has been calculated with observations taken with the $H2$ filter of IRDIS.
For July data, the ambient temperature during the observations was on average T$_{amb}$ = 14.9$^{\circ}$C, which means that the expected pixel scale is $12.255 \pm0.012$, compatible with the value listed in Table~\ref{t:plate}. We use this corrected value of the plate scale, which accounts for the temperature during the observations. The pixel scale of the filter $H3$ is $12.250 \pm0.012$. }

\subsection{Broad and dual-band imaging}
\label{sec:irdis_red}
The IRDIS data were preprocessed in the following way: first we subtracted from each datacube the corresponding sky background taken right after the sequence, we divided by flat-field images, and we applied a procedure to identify and interpolate  bad-pixels. The frames were recentered using the initial star center exposure with the four satellite spots. The master backgrounds and master flat-field have been obtained with DRH software.

The second step of the reduction has been carried out with four independent pipelines performing ADI. We did not employ spectral differential imaging techniques  \citep[SDI;][]{1999PASP..111..587R} between the IRDIS filter pairs to avoid a flux loss that is not straightforward because of self-subtraction \citep{Maire14}. The first data reduction method is based on the principal component analysis (PCA) technique that performs the Karhunen--Lo{\`e}ve Image Projection (KLIP) algorithm, following the model of \citet{2012ApJ...755L..28S}. This pipeline has been presented in \citet{2014A&A...572A..85Z}. For each product of the preprocessing phase, we created a PCA library based on all the other collapsed datacubes of the sequence, and we derotated the product of the PCA as a function of the parallactic angle of the reference datacube.  A final image was obtained as the median of all the rotated PCA images. In some cases, the data was binned temporally to reduce the number of frames and increase the analysis speed. 

The second pipeline is a more recent version of the T-LOCI algorithm presented in \citet{marois14}.  For each dual-band filter sequence, we calibrate the speckle pattern of each individual frame, rotate the frames to align north up, and median combined all the frames to obtain the final image. 

To perform the reduction of the images in $J$ band, we exploited a third pipeline based on the method of the radial ADI (rADI) as illustrated in \citet{2012A&A...542A..41C}. 

Finally, the BB\_$H$ data were reduced with a fourth independent pipeline: datacubes were high-pass filtered by subtracting from each image a version of the image smoothed by a median filter with a large 11-pixel-wide box width before entering a PCA algorithm.

\begin{figure*}
\begin{center}
\includegraphics[width=\textwidth]{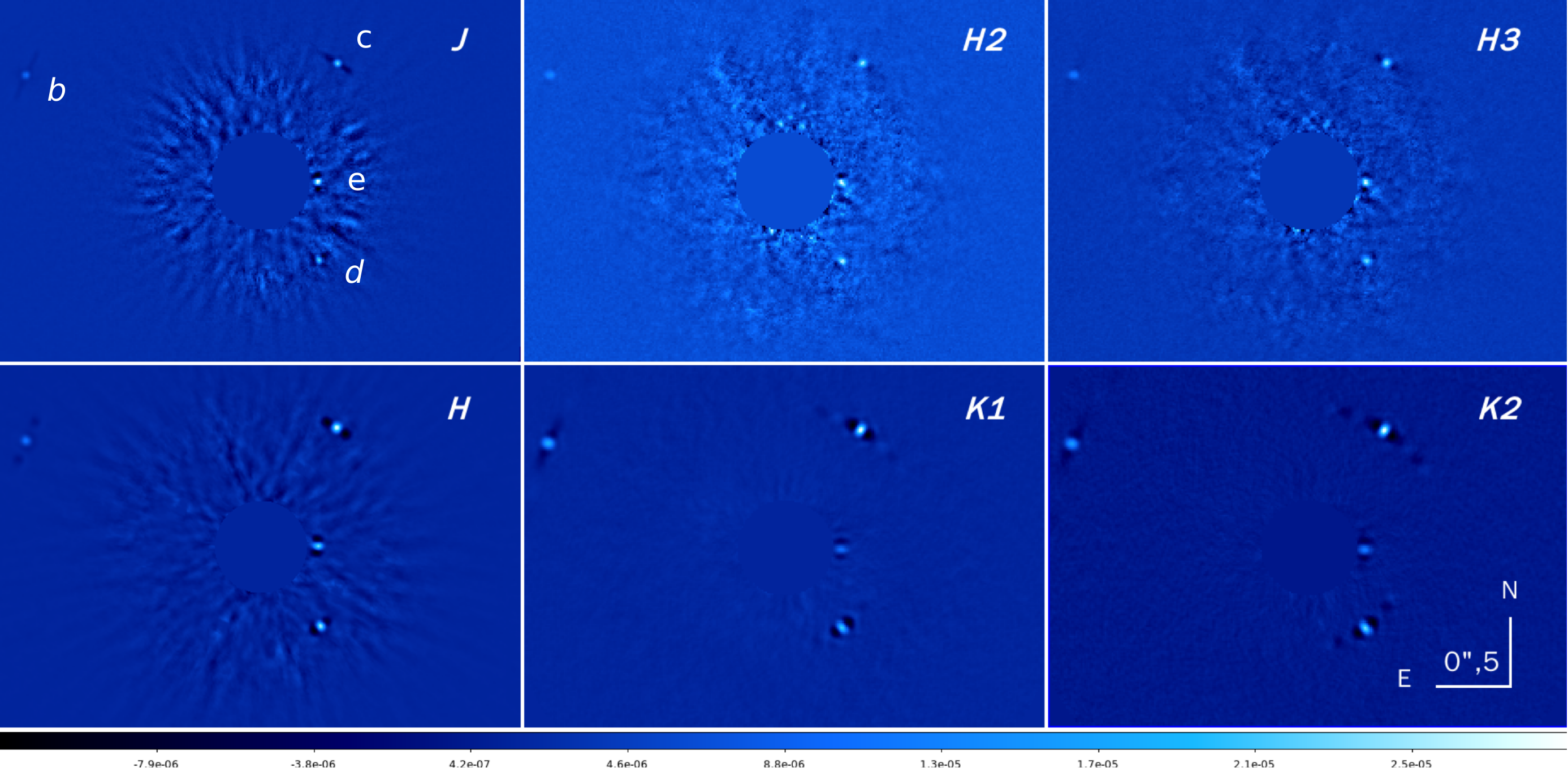}
\caption{IRDIS images for several filters with the four planets around HR\,8799. The color scale is the same for all images, in unit of contrast with respect to the host star. }
\label{f:k2}
\end{center}
\end{figure*}

The final images obtained for all the IRDIS filters are shown in Fig.~\ref{f:k2}. No significant detection {of a candidate planet} is found closer than planet e in any of the images. The signal-to-noise ratio (SNR) for the planets goes from a minimum of 11 (planet d in $J$ band) to a maximum of 220 (planet c in $K1$ band).  Section~6.1 of \citet{2014A&A...572A..85Z} gives a detailed explanation of how we calculated the SNR.

To measure photometry and astrometry, we exploited the method of the ``fake negative planets'' \citep{2010Sci...329...57L,2010SPIE.7736E..1JM,2011A&A...528L..15B}, as illustrated in \citet{2014A&A...572A..85Z} and \citet{galicher11}. To build models of the planet images accounting for the ADI self-subtraction, we used the median of the unsaturated images of HR\,8799 taken right before and after the sequences and the previously determined LOCI coefficients or PCA eigenvectors. No unsaturated images of HR\,8799 were recorded during science verification, so HR\,8799\,b served as a PSF to build the planetary system model. We then adjusted the models to the real planet images to estimate the astrometry and photometry of the planets.  We obtained the final results for each independent pipeline with a minimum $\chi^{2}$ fitting, where we simultaneously varied the flux and  position of the fake negative planets \citep[as described in][]{2014A&A...572A..85Z}. The error bars for each reduction account for the flux variations of the unsaturated images of HR\,8799, the fluctuations of the results as a function of the PCA/LOCI parameters, and the accuracy of the fitting of the planet image models to the real images.

Results from each pipeline have been combined together excluding single results, which deviate more than the standard deviation, $\sigma$, of the values themselves. The error bars have been calculated as the quadratic sum of each independent reduction error bar plus the standard deviation of the values themselves.  The contrast in magnitude obtained for the four planets is listed in Table~\ref{t:phot}, while their astrometric positions are shown in Table~\ref{t:astro}. {The error budget is listed in Table~\ref{t:errastro}.}

\begin{table*}
\caption{IRDIS contrast in magnitudes and absolute magnitudes for HR\,8799 bcde. The broad-band $H$ magnitudes are not provided as no off-axis PFS was taken during the sequence. This sequence is used for the astrometric analysis only in this paper. We refer to \citet{2016ApJ...820...40A} for a detailed discussion on the possible variability of these planets.  } 
\label{t:phot}
\renewcommand{\footnoterule}{}  
\centering
\begin{tabular}{ccccc}
\hline
\hline
           Filter           &  b   & c & d & e   \\
\hline
\multicolumn{5}{c}{\it Contrast [mag]}\\
\hline
$\Delta J$       & $14.39\pm0.09$   & $13.21\pm0.13$ & $13.20\pm0.37$  & $13.01\pm0.21$ \\
                  
$\Delta H2$      & $12.80\pm0.14$  & $11.81\pm0.12$ & $11.74\pm0.17$ & $11.63\pm0.20$ \\
                  
$\Delta H3$      & $12.50\pm0.10$ & $11.50\pm0.10$ & $11.57\pm0.16$  & $11.40\pm0.21$ \\
                  
$\Delta K1$      & $11.91\pm0.06$ & $10.95\pm0.05$ & $10.96\pm0.07$ & $10.88\pm0.10$\\
                  
$\Delta K2$      & $11.73\pm0.09$ & $10.62\pm0.07$ & $10.60\pm0.10$ & $10.58\pm0.11$\\
\hline
\multicolumn{5}{c}{\it Absolute magnitude [mag]}\\
\hline
$J$       &$16.80\pm    0.09$ & $15.62\pm       0.13$  &$       15.61\pm        0.37$  &$     15.42\pm        0.21$ \\
                                                                                                
$ H2$      &$15.10\pm   0.14$  &$14.11\pm       0.12$  &$       14.04\pm        0.17$  &$     13.93\pm        0.20$ \\
                                                                                                
$ H3$      &$14.80\pm   0.10$  &$13.80\pm       0.10$  &$       13.87\pm        0.16$  &$     13.70\pm        0.21$ \\
                                                                                                
$ K1$      &$14.17\pm   0.06$  &$13.21\pm       0.05$  &$       13.22\pm        0.07$  &$     13.14\pm        0.10$ \\
                                                                                                
$ K2$      &$13.99\pm   0.09$  &$12.88\pm       0.07$  &$       12.86\pm        0.10$  &$     12.84\pm        0.11$ \\
\hline

\end{tabular}
\end{table*}

\begin{table}
\caption{Astrometric positions of the planets around HR\,8799 in date 2014.53, derived from IRDIS filters $H2H3$, and measurements in date 2014.62, from IFS $YH$ band. The planet positions in broad $H$-band in date 2014.93 are also shown.} 
\label{t:astro}
\centering
\begin{tabular}{llrr}
\hline
\hline
 Epoch &          Planet         &  $\Delta$RA (\as)   &  $\Delta$Dec (\as)  \\
\hline
2014.53 &HR\,8799\,b      & $1.570\pm0.006$  &  $0.707\pm0.006$ \\
&HR\,8799\,c    &  $-0.522\pm0.004$     & $0.791\pm0.004$  \\
&HR\,8799\,d      & $-0.390\pm0.005$ &  $-0.530\pm0.006$   \\
&HR\,8799\,e   & $-0.386\pm0.009$       & $-0.008\pm0.009$   \\
\hline
2014.62 &HR\,8799\,d      & $-0.391\pm0.004$  & $-0.529\pm0.004$ \\
&HR\,8799\,e   &   $-0.384\pm0.002$     & $-0.005\pm0.002$\\
\hline
2014.93& HR\,8799\,b      & $1.574\pm0.005$  &  $0.703\pm0.005$ \\
&HR\,8799\,c    &  $-0.518\pm0.004$     & $0.797\pm0.004$  \\
&HR\,8799\,d      & $-0.402\pm0.004$ &  $-0.523\pm0.004$   \\
&HR\,8799\,e   & $-0.384\pm0.010$       & $0.014\pm0.010$   \\
\hline
\end{tabular}
\end{table}

\begin{table}
\caption{{Astrometric error budget for both IRDIS and IFS measurements. For each planet the error was calculated as a combination of: plate scale uncertainty ($\sigma_{PS}$); dithering procedure accuracy ($\sigma_{Dith} = 0.74$ mas, here approximated to 1 mas); star center position uncertainty ($\sigma_{SC}$) { of 1.2 mas, derived from observation of bright stars during commissioning runs}; planet position uncertainty ($\sigma_{PP}$). For the sake of simplicity we listed just the error along the x axis of the detector for the planet position measurement. The errors due to the anamorphism uncertainty and the true north uncertainty are less than 0.5 mas for any planet, so they are not listed here, but included in the calculation of the error budget.  }} 
\label{t:errastro}
\centering
\begin{tabular}{lccccc}
\hline
\hline
Epoch &           Planet     &  $\sigma_{PS}$ (\as) & $\sigma_{Dith}$ (\as)& $\sigma_{SC}$ (\as) & $\sigma_{PP}$ (\as)  \\
\hline
2014.53 &b       & 0.002&0.001  &0.001  &  0.005         \\
&c      &0.001&0.001 &0.001 & 0.004  \\
&d      &0.001&0.001            &0.001 &  0.005    \\
&e   & 0.000&0.001      &0.001 & 0.009             \\
\hline
2014.62 &d      &0.001&-                &0.001 &  0.004    \\
&e   & 0.001&-  &0.001 & 0.001             \\
\hline
2014.93& b       & 0.002 &0.001 &0.001  &  0.004         \\
&c      &0.001&0.001 &0.001 & 0.004  \\
&d      &0.001&0.001            &0.001 &  0.004    \\
&e   & 0.000&0.001      &0.001 & 0.010             \\
\hline
\end{tabular}
\end{table}

\subsection{IFS data reduction and spectra extraction}
\label{sec:ifs_red}
We obtained IFS data in both $YJ$ (0.95--1.35~$\mu$m, spectral resolution $R$\,$\sim$\,54) and $YH$ (0.95--1.65~$\mu$m, $R$\,$\sim$\,33) modes \citep{2008SPIE.7014E..3EC}. We only present the $YH$ data  because of the poor quality of the $YJ$ data obtained in July 2014. The IFS raw data were preprocessed using the DRH software, up to the creation of the calibrated spectral datacubes.

The preprocessing consists of background subtraction and flat-field calibration.  Following this, 
the locations of the spectral traces are determined using a calibration image where the integral field unit (IFU) is uniformly 
illuminated with a white lamp in such a way that the detector is completely filled with spectra.  For wavelength calibration, the IFU is illuminated with four monochromatic lasers of known wavelength. An instrument flat is used to correct for the different response of the lenslets to a uniform illumination and detector quantum efficiency variations.  After this last procedure, we are left with calibrated spectral datacubes comprised of 39 monochromatic images. 

An additional step using custom IDL tools has been introduced for better handling of bad pixels and to implement spectral cross-talk corrections \citep{2015A&A...576A.121M}. After the creation of the calibrated spectral datacubes, the differential imaging part of the data reduction is performed applying different pipelines written in IDL.
\begin{figure}
\begin{center}
\includegraphics[width=0.40\textwidth]{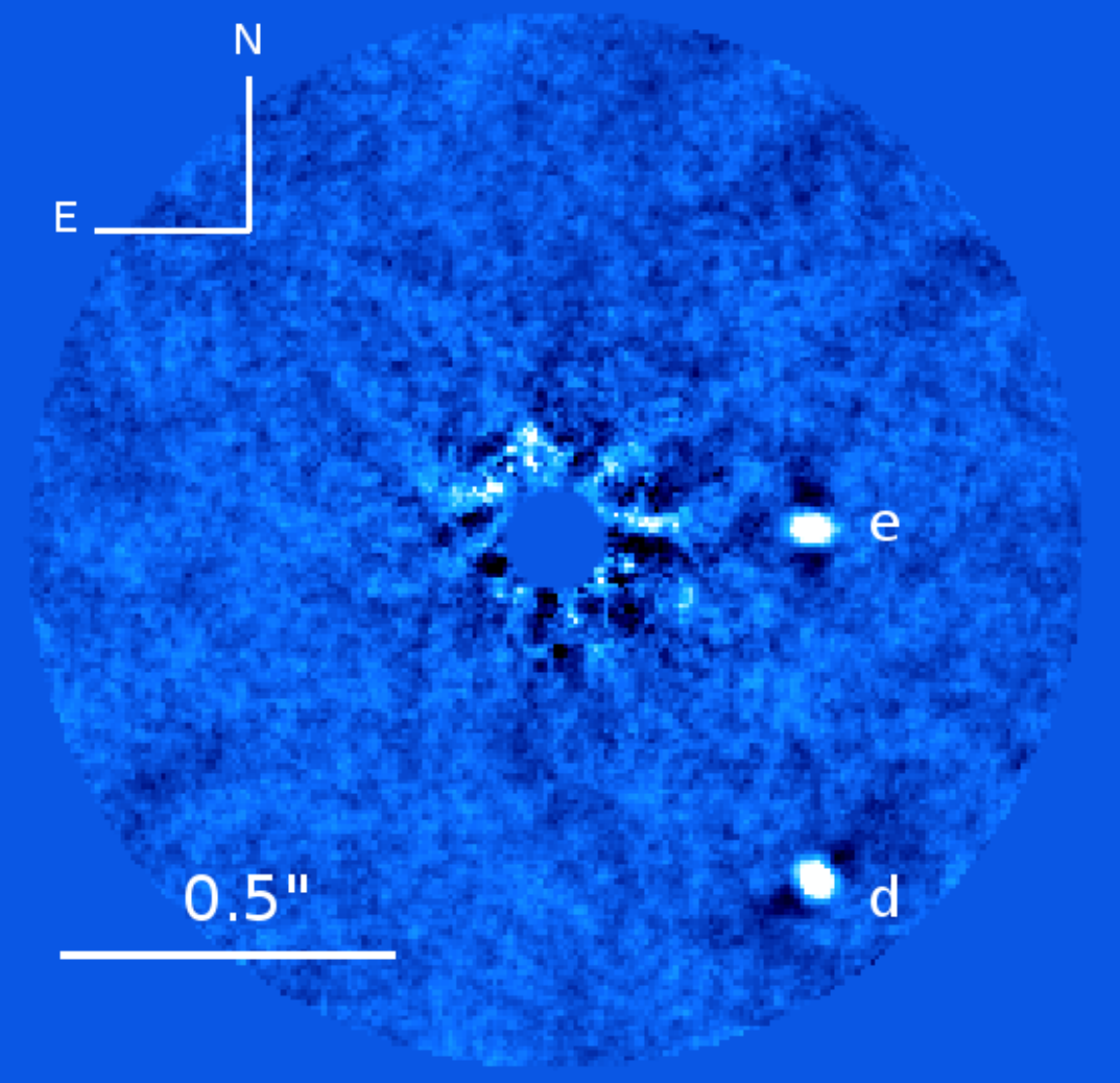}
\caption{Image of the 38th IFS spectral channel (1.63~$\mu$m) with planets HR\,8799\,de obtained with KLIP reduction.}
\label{f:ifs}
\end{center}
\end{figure}

We have adopted two different approaches to searching for additional companions in the HR\,8799 system as well as extracting spectra of the detected planets. Both approaches provided detection of planets d and e with a median SNR along the IFS channels of $\sim$ 20. 
To search for additional planets, we utilize the KLIP algorithm, exploiting all the frames at
different wavelengths and at different angular positions. No significant signal has been found closer than planet e. 

For extraction of spectra of known objects, we first utilize ADI separately for each individual IFS channel. As shown in Table~\ref{t:obs}, for the first $YH$ observations we took 16 datacubes comprised of three frames with a DIT of 60 s, while for the second observations we took 16 datacubes comprised of two frames with a DIT of 100 s. We decided to use all 80 frames without performing any binning in time, rescaling the frames of the second data set to the total integration time of the first.  For each spectral frame we 
performed PCA fitting using  the frames of the same channel of the other datacubes as a library. 
To avoid self-subtraction, we selected the frames according to the planet position in each 
image of the library, and rejected the frames where the planet centroid moved less than 
$\sim$0.5$\lambda/D$ with respect to the reference frame, following the example presented in \citet{2015ApJ...803...31P}. For each planet the library is composed of a different set of frames, however, we also attempted data reductions without performing any selection of the frames. In this case, we used a small number of modes to limit the biases of the PCA.   
We used different number of modes  ($<$ 80), but the result converged rapidly after a few modes (between 2 and 5). 
To retrieve the signal of a planet, we tested the impact of our reductions with synthetic planets injected in the preprocessed datacubes 
at different positions or in the forward modeling \citep[see description in][]{2012ApJ...755L..28S}, assuming the same PCA modes and aperture as employed for the spectral extraction in the science data. For the forward modeling, we used the off-axis PSF obtained after the observing sequence because of a slight saturation of the off-axis PSF obtained before the observation.  
An example of the output of this technique is shown in Fig~\ref{f:ifs}. 

A second approach, where ADI and SDI were used together, was then implemented.  The single exposures in each of the 32 spectral datacubes were averaged. The PSF library used for the PCA fitting taking all frames into account, both along the temporal and spectral dimensions, where the PSF has moved by more than a separation criteria $N_{\delta}$. In this scheme, the library is constructed using the frames spatially rescaled according to their respective wavelength. Various tests showed that a value of $N_{\delta}=0.8$~FWHM gives good results in terms of SNR for the detection of the planets. To account for the flux losses induced by the speckle subtraction algorithm, forward modeling is performed via the off-axis PSF of the  above-mentioned star as a model for the PSF of the planet. The model is projected on the same PCA modes as the science data, and the flux loss is measured as the attenuation of the flux for the model before and after projection on the PCA modes. The fluxes of the planets are measured in the final spectral channels with aperture photometry in an aperture of diameter 0.8~$\lambda/D$ ($\sim$ 4 pixels). The flux is finally corrected for the flux loss estimated using the forward modeling. The error bars on the photometry were measured in the same way as for the IRDIS data described in Sect.~\ref{sec:irdis_red}, except that we did not consider the flux variation during the exposure, as one of the two PSFs is slightly saturated.

With this approach, we obtained eight spectral extractions from four independent pipelines. For each planet, a final estimate of the flux ratio between the star and  planet was obtained by computing the median of the results.

Astrometric measurements for planets HR\,8799\,de measured from the IFS are shown in Table~\ref{t:astro}. { The error budget is presented in Table~\ref{t:errastro}.  The error bars are compatible with the expected ones for these targets, following the simulations of \citet{2014A&A...572A..85Z}.} 

\begin{figure}
\begin{center}
\includegraphics[width=0.48\textwidth]{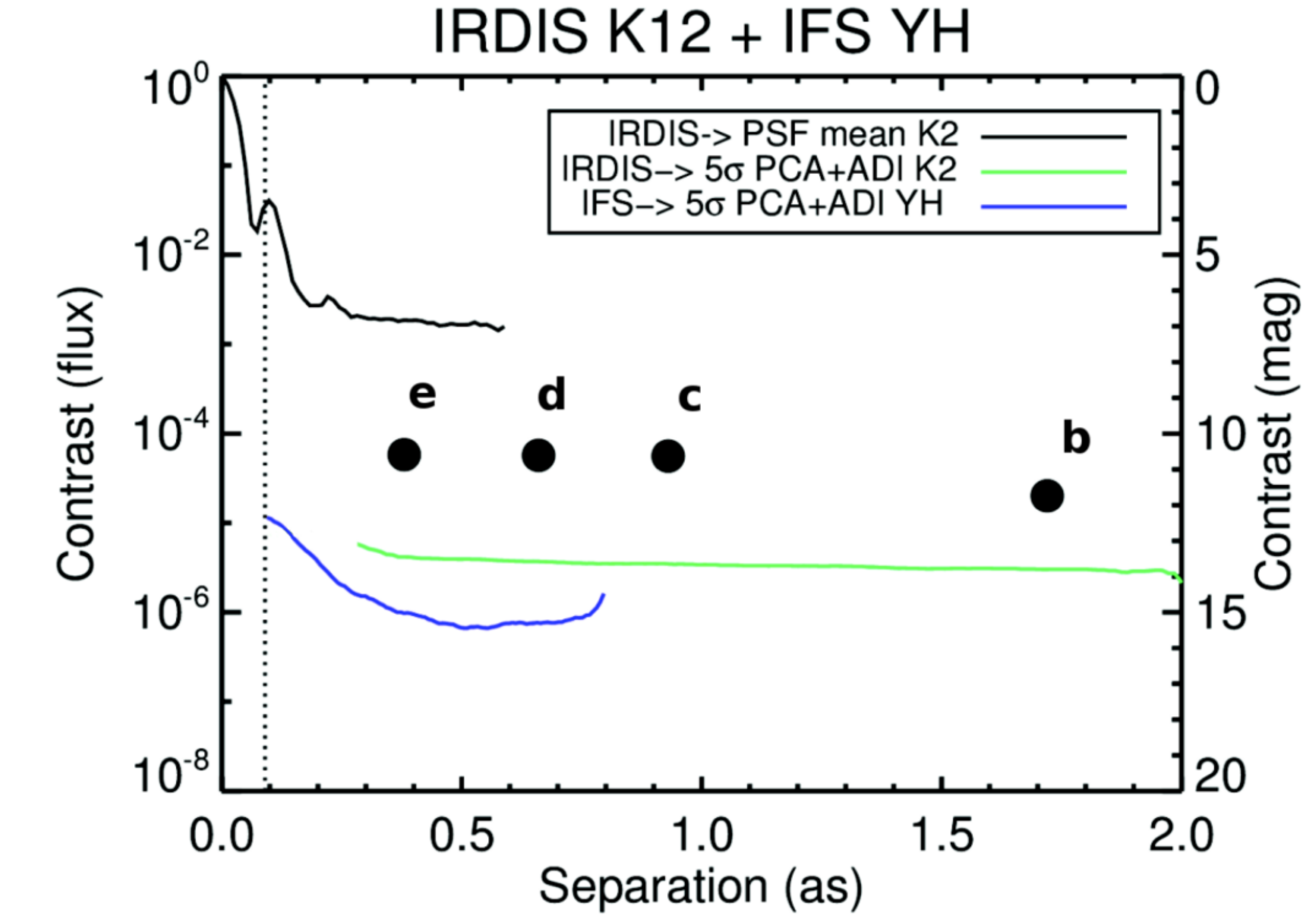}
\caption{Contrast { and off-axis PSFs} curves for the IRDIFS\_EXT data set. The 5$\sigma$ contrast for the two instruments are plotted as a function of the angular separation. The off-axis PSF of the star is represented as a black continuous line. Planets around HR\,8799 are also shown as photometric $K2$ points, the error bars are inside the dimension of the dots. The dotted vertical line indicates the coronagraph IWA.}
\label{f:contrast}
\end{center}
\end{figure}

The 5$\sigma$ contrast curves of IRDIS and IFS for the IRDIFS\_EXT data set are shown in Fig.~\ref{f:contrast}. The limits were calculated by measuring the standard deviation of the residuals in annuli of width
1 $\lambda$/D at increasing angular separation, normalized for the flux of the corresponding PSF.

Contrast in the $K2$ band of the planets around HR\,8799 are overplotted. IRDIS contrast curve has been calculated on the KLIP product of the filter $K2$, while the IFS contrast curve has been obtained on the KLIP result with just ADI ($YH$ band). 

\begin{figure}
\begin{center}
\includegraphics[width=0.48\textwidth]{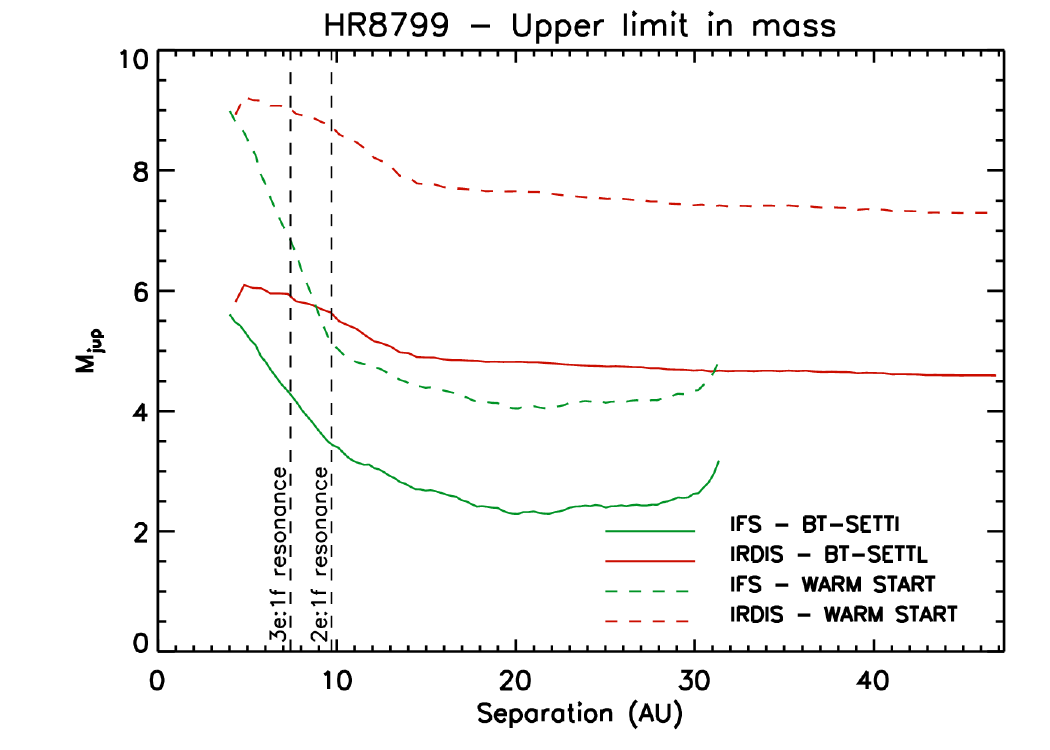}
\caption{Mass limits deduced from the contrast curves for the IRDIS and IFS instruments. An age of 30 Myr is assumed. { Two} different models are represented: the ``hot-start models'' BT-SETTL models, and the ``warm-start models'' from \citet{2012ApJ...745..174S}. The separations where a possible planet f is expected to be \citep[from][]{2014MNRAS.440.3140G} are shown as dashed black lines.}
\label{f:mass_limit}
\end{center}
\end{figure}

The mass limits deduced from the contrast curves of Fig.~\ref{f:contrast} are shown in Fig.~\ref{f:mass_limit}. To calculate the mass limits, we assumed an age of 30 Myr for the system and two different models: the BT-SETTL \citep{2012EAS....57....3A} for ``hot-start models'' as well as the ``warm-start models'' from \citet{2012ApJ...745..174S}.

 Following \citet{2014MNRAS.440.3140G}, a possible planet could be located at the 3e:1f and 2e:1f resonance orbits, corresponding to separation of $\sim 7.5$ and $\sim 10$ AU. The two separations are overplotted in Fig.~\ref{f:mass_limit}. In their analysis they found that a planet of mass 2-8 \MJup in the 3e:1f orbit and 1.5-5 \MJup at the 2e:1f orbit could be present without perturbations on the stability of the system. We found that with IFS we are able to detect planets with mass of 4.5-7 \MJup at the separation of the 3e:1f resonance and 3-5 \MJup at the resonance 2e:1f. { This means that  2-4.5 \MJup planets are not excluded by our estimations for the 3e:1f orbit, as  we do not exclude 1.5-3 \MJup planets for the 2e:1f
orbit.}

\section{Astrometric fit of the orbits}
\label{Sec:astro}

{ In this section, we present an illustration of the accuracy of SPHERE in measuring the astrometric positions of low-mass companions with respect to the previous generation of high-contrast imagers. We performed an astrometric fit for the orbits of planets HR\,8799\,bcde.
The aim of this section is to present sample orbital solutions compatible with the astrometric data, but a detailed orbital analysis is beyond the scope of this paper.}

We combined the astrometric positions available from the literature, including \citet{2008Sci...322.1348M, 2009ApJ...694L.148L, 2009ApJ...696L...1F, 2010ApJ...716..417H, 2010Natur.468.1080M, 2011ApJ...729..128C,2011A&A...528A.134B, 2011ApJ...739L..41G,2011ApJ...741...55S, 2012ApJ...755L..34C,2013A&A...549A..52E,2014ApJ...795..133C,2015A&A...576A.133M, 2015ApJ...803...31P} with our SPHERE data (Table~\ref{t:astro}). {We rely on the nominal error bars from the individual
references, although systematic errors between different instruments and
analysis procedures may be present. The SPHERE astrometric measurements
are compatible with the literature data and achieve an accuracy similar or
better with respect to the published measurements, which is a good test
for this new instrument.}

{The orbital fit is a minimum $\chi^2$ fitting, as described in \citet{2013A&A...549A..52E}, where all the four orbits are simultaneously taken into consideration, using the assumption that the planet periods are commensurate.  The IFS measurement is used for the closest
planet e, as it provides smaller error bars. We assumed circular and non-coplanar orbits and tested three different configurations for the orbital period ratios between the planets.}

\begin{figure*}
\begin{center}
\includegraphics[width=\textwidth]{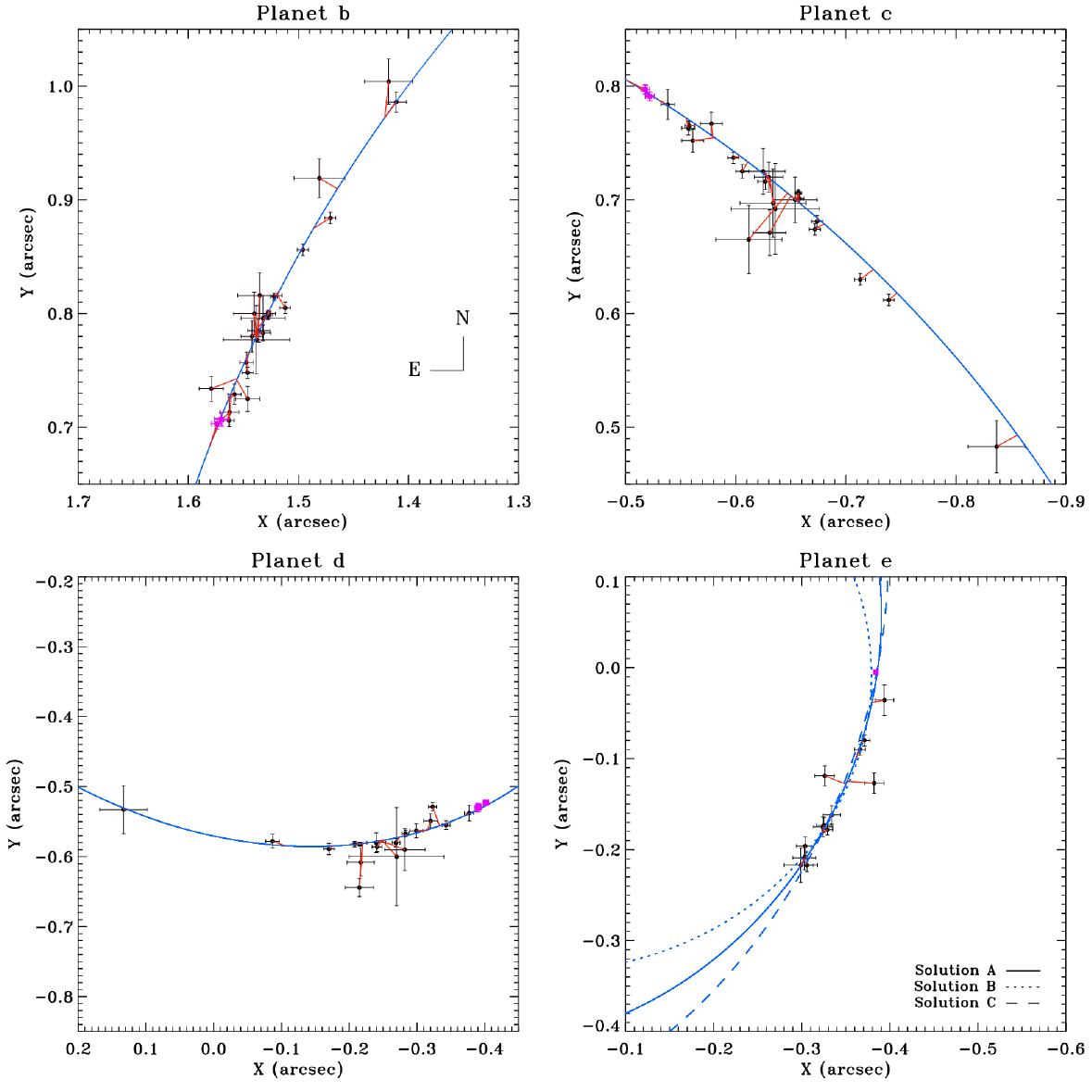}
\caption{{Relative astrometry for the planets around HR\,8799. The points available from the literature are represented with black filled circles, while new SPHERE measurements are represented with purple filled squares. The solid blue lines represent the orbital solution labeled A in Table~\ref{t:orbits}. Red lines connect the predicted and observed positions for all the data points. The dotted and dashed lines in the panel for HR 8799 e indicate the orbital solutions labeled B and C in Table 5, respectively.}}
\label{f:plot_astro}
\end{center}
\end{figure*}

{The fitted orbits are shown in Fig.~\ref{f:plot_astro}. The parameters of the orbital solutions are summarized in Table~\ref{t:orbits}. The uncertainties on the parameters were estimated using a quadratic combination of statistical and systematic errors. The statistical errors stem from the sensitivity of the fit to the errors bars on the astrometric measurements and are derived from 1000 random trials of the astrometric measurements assuming normal distributions \citep{2015A&A...576A.133M}. The systematic errors are related to the uncertainties on the stellar mass and distance (Table~\ref{t:orbits}), which propagate into the derivation of the semimajor axis. To assess them, we performed two sets of orbital fits. The first set assumes the nominal values for the stellar mass and distance, and the second set assumes a combination of values at the edge of the derived ranges, which provides the largest offset in orbital periods. Two orbital solutions are favored with similar reduced $\chi^2$, with the orbital period ratios 2d:1e and 3d:1e. The IFS data point for planet e is out of the orbital solution with the orbital period ratio 5d:2e at $\sim$2.5$\sigma$ significance level. Further astrometric monitoring is required to distinguish between the orbital period ratios 2d:1e and 3d:1e.}

\begin{table}
\caption{{{Orbital elements fitted on the astrometric data in Fig.~\ref{f:plot_astro}.}}}
\label{t:orbits}
\begin{center}
\begin{tabular}{l c c c}
\hline\hline
Parameter & A & B & C \\
\hline
HR~8799 b & & & \\
$P$ (yr) & 455.88$\pm$\,24.70 & 455.88$\pm$\,24.70 & 455.88$\pm$\,24.70 \\
$i$ ($^{\circ}$) & 30.27\,$\pm$\,4.95 & 30.27\,$\pm$\,4.95 & 30.27\,$\pm$\,4.95 \\
$\Omega$ ($^{\circ}$) & 60.89\,$\pm$\,4.40 & 60.89\,$\pm$\,4.40 & 60.89\,$\pm$\,4.40 \\
$e$ & -- & -- & -- \\
$\omega$ ($^{\circ}$) & -- & -- & -- \\
$T0$ (yr) & 2006.55\,$\pm$\,6.44 & 2006.55\,$\pm$\,6.44 & 2006.55\,$\pm$\,6.44 \\
$a$ (AU) & 67.96\,$\pm$\,1.85 & 67.96\,$\pm$\,1.85 & 67.96\,$\pm$\,1.85 \\
\hline
HR~8799 c & & & \\
$P$ (yr) & 227.94\,$\pm$\,12.35 & 227.94\,$\pm$\,12.35 & 227.94\,$\pm$\,12.35 \\
$i$ ($^{\circ}$) & 29.43\,$\pm$\,0.35 & 29.43\,$\pm$\,0.35 & 29.43\,$\pm$\,0.35 \\
$\Omega$ ($^{\circ}$) & 65.68\,$\pm$\,2.67 & 65.68\,$\pm$\,2.67 & 65.68\,$\pm$\,2.67 \\
$e$ & -- & -- & -- \\
$\omega$ ($^{\circ}$) & -- & -- & -- \\
$T0$ (yr) & 1848.18\,$\pm$\,9.84 & 1848.18\,$\pm$\,9.84 & 1848.18\,$\pm$\,9.84 \\
$a$ (AU) & 42.81\,$\pm$\,1.16 & 42.81\,$\pm$\,1.16 & 42.81\,$\pm$\,1.16 \\
\hline
HR~8799 d & & & \\
$P$ (yr) & 113.97\,$\pm$\,6.18 & 113.97\,$\pm$\,6.18 & 113.97\,$\pm$\,6.18 \\
$i$ ($^{\circ}$) & 38.63\,$\pm$\,2.84 & 38.63\,$\pm$\,2.84 & 38.63\,$\pm$\,2.84 \\
$\Omega$ ($^{\circ}$) & 56.09\,$\pm$\,3.78 & 56.09\,$\pm$\,3.78 & 56.09\,$\pm$\,3.78 \\
$e$ & -- & -- & -- \\
$\omega$ ($^{\circ}$) & -- & -- & -- \\
$T0$ (yr) & 1965.27\,$\pm$\,1.55 & 1965.27\,$\pm$\,1.55 & 1965.27\,$\pm$\,1.55 \\
$a$ (AU) & 26.97\,$\pm$\,0.73 & 26.97\,$\pm$\,0.73 & 26.97\,$\pm$\,0.73 \\
\hline
HR~8799 e & & & \\
$P$ (yr) & 56.99\,$\pm$\,3.09 & 46.99\,$\pm$\,2.79 & 75.19\,$\pm$\,4.05 \\
$i$ ($^{\circ}$) & 30.95\,$\pm$\,1.43 & 28.43\,$\pm$\,5.39 & 44.51\,$\pm$\,0.45 \\
$\Omega$ ($^{\circ}$) & 145.73\,$\pm$\,7.11 & 85.50\,$\pm$\,3.47 & 156.08\,$\pm$\,2.54 \\
$e$ & -- & -- & -- \\
$\omega$ ($^{\circ}$) & -- & -- & -- \\
$T0$ (yr) & 1995.61\,$\pm$\,0.93 & 1990.54\,$\pm$\,1.02 & 1992.31\,$\pm$\,1.17 \\
$a$ (AU) & 16.99\,$\pm$\,0.46 & 14.94\,$\pm$\,0.46 & 20.44\,$\pm$\,0.55 \\
\hline
$\chi_{\mathrm{red,b}}^2$ & 2.33 & 2.33 & 2.33 \\
$\chi_{\mathrm{red,c}}^2$ & 2.04 & 2.04 & 2.04 \\
$\chi_{\mathrm{red,d}}^2$ & 1.92 & 1.92 & 1.92 \\
$\chi_{\mathrm{red,e}}^2$ & 1.21 & 3.44 & 1.18 \\
\hline
$M_{\mathrm{star}}$ $(M_{\odot})$ & 1.51$^{+0.038}_{-0.024}$ & 1.51$^{+0.038}_{-0.024}$ & 1.51$^{+0.038}_{-0.024}$ \\
$d$ (pc) & 39.4\,$\pm$\,1.1 & 39.4\,$\pm$\,1.1 & 39.4\,$\pm$\,1.1 \\
\hline
\end{tabular}
\end{center}
\tablefoot{{{Case A: orbital solution setting circular orbits and 8b:4c:2d:1e mean motion resonance. Case B: orbital solution setting circular orbits and 4b:2c:1d and 5d:2e mean motion resonances. Case C: orbital solution setting circular orbits and 4b:2c:1d and 3d:2e mean motion resonances.\\
For each planet, the notations refer to the orbital period, inclination, longitude of the ascending node, eccentricity, argument of periapsis, time of periapsis passage, and semimajor axis. {The error bars on the orbital parameters are at 1~$\sigma$ (see text).} We also indicate the reduced $\chi^2$ {for the nominal orbital solution for each planet,} and the mass and distance ranges assumed for the host star for the orbital fitting \citep{2012ApJ...761...57B, 2007A&A...474..653V}. The semimajor axes are derived using Kepler's third law assuming the fitted orbital periods and that the planet masses are negligible with respect to the mass of the host star.}}} \\
\end{table}

\section{Comparison to known objects}
\label{sec:specphotanalysis}
In this section, we used the extracted photometry from IRSDIS and spectrophotometry from the IFS to explore characterization of the SEDs of the planetary mass objects surrounding HR 8799. An in-depth study of these properties is presented in a companion paper (Bonnefoy et al., {A\&A accepted}).
\subsection{Colors}

We first compare the IRDIS colors of HR\,8799\,bcde to those of MLT field dwarfs.  The colors of the field dwarfs were synthetized from SpeXPrism spectral library\footnote{http://pono.ucsd.edu/$\sim$adam/browndwarfs/spexprism/}. For that purpose, we smoothed the low-resolution spectra of the sources to the resolution of the narrowest IRDIS filter, using the IRDIS filter passbands, a model of the Paranal atmospheric transmission generated with the ESO \texttt{Skycalc} web application\footnote{http://www.eso.org/obser\-ving/etc/bin/gen/form?INS.MO\-DE=swspectr+INS.NA\-ME=SKYCALC} \citep{2012A&A...543A..92N, 2013A&A...560A..91J}, and a model spectrum of Vega \citep{2007ASPC..364..315B}.

\begin{figure}
\begin{center}
\includegraphics[width=\columnwidth]{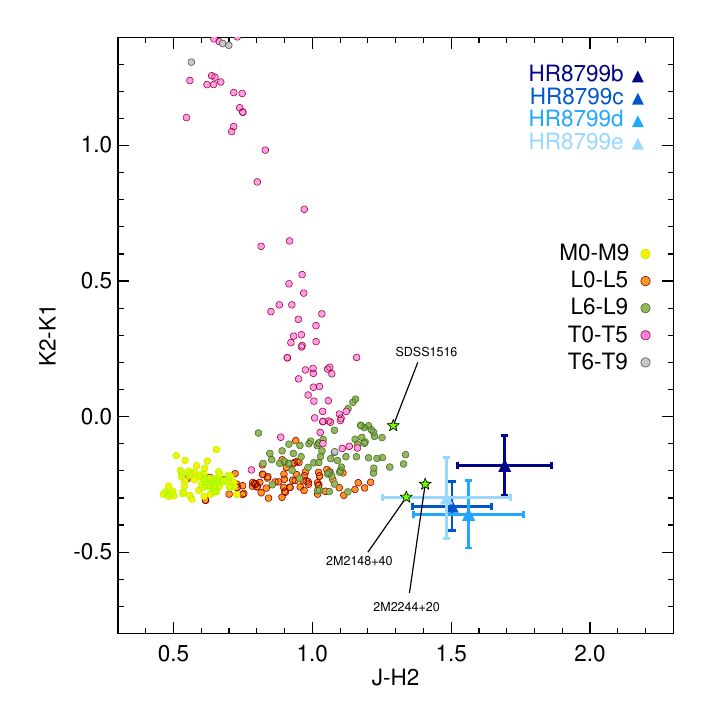}
\caption{Comparison of HR\,8799\,bcde colors based on IRDIS photometry to those of M, L, T field dwarfs (dots).}
\label{f:JH2K2K1}
\end{center}
\end{figure}

\begin{figure}
\begin{center}
\includegraphics[width=\columnwidth]{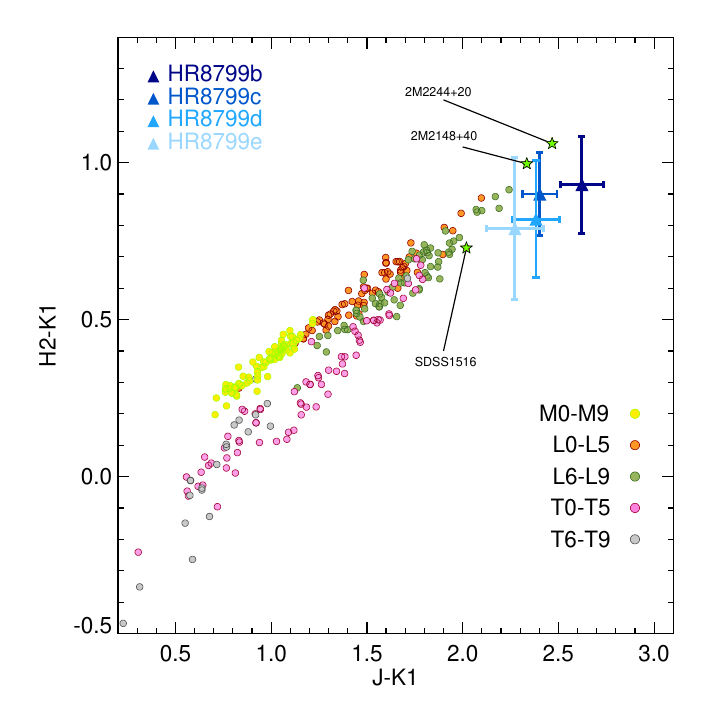}
\caption{Same as Fig.~\ref{f:JH2K2K1}, except for the $J-K$1 versus $H2-K1$ colors.}
\label{f:JK1H2K1}
\end{center}
\end{figure}

Certain two-color diagrams enable discrimination of field dwarf objects with different spectral types at the L-T transition. They are shown in Figures~\ref{f:JH2K2K1} and \ref{f:JK1H2K1}. The planets are located at the transition between late-L and early-T dwarfs, but have redder colors than the dwarf objects (i.e. are brighter in the K-band than expected). This is in agreement with the location of the objects in color-magnitude diagrams \citep{2008Sci...322.1348M, 2011ApJ...729..128C}. 

\subsection{Near-infrared slope}
 \label{Sec:AppendixA}

\begin{figure}
\begin{center}
\includegraphics[width=0.45\textwidth]{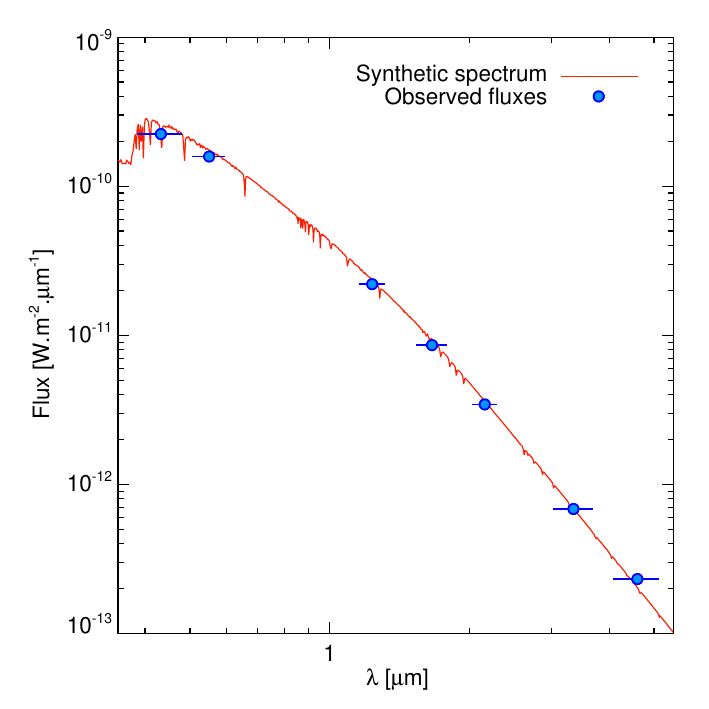}
\caption{Synthetic ATLAS9 spectrum adjusted to the apparent fluxes of HR\,8799\,A.}
\label{FigApA:Fig1}
\end{center}
\end{figure}

 We converted the flux-ratio (IFS spectra, IRDIS photometry) between the HR\,8799 planets and their host star into apparent flux, corrected from telluric absorptions with a model spectrum of the star \citep[ATLAS9 Model grid;][]{2004astro.ph..5087C} with $T_{eff}=7500$ K, log g=4.5, M/H=0, $v_{turb}=0$ km/s scaled in flux to fit the Tycho2 $BV$, 2MASS $JHK_{S}$, and WISE $W1W2$ photometry \citep{2000A&A...355L..27H, 2003tmc..book.....C, 2012yCat.2311....0C}. This synthetic spectrum was found to provide a good fit for the existing photometry. The scaled spectrum of the star is shown in Fig.~\ref{FigApA:Fig1}.

\begin{figure*}
\begin{center}
\includegraphics[width=1.3\columnwidth]{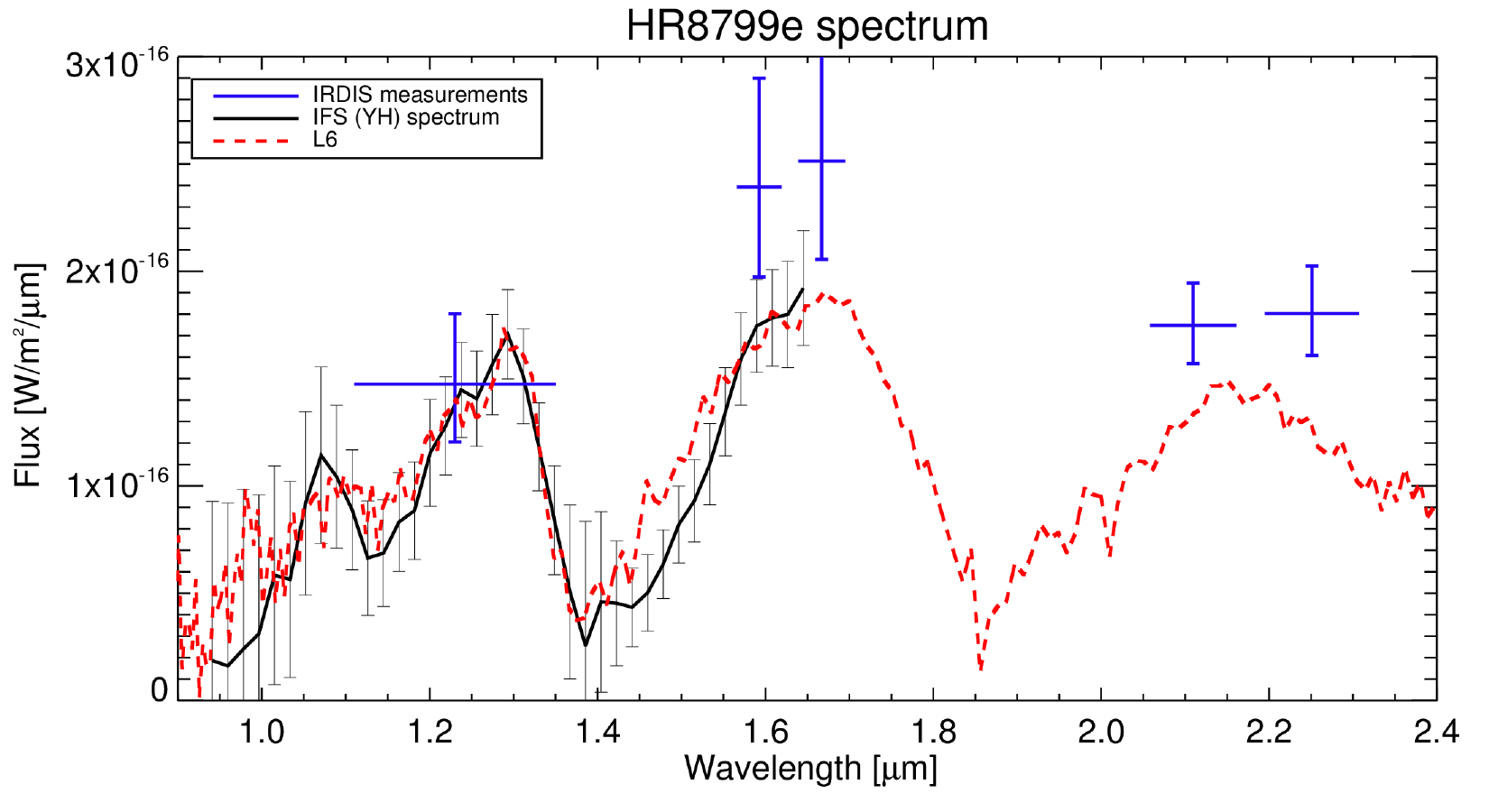}
\caption{Spectrum of planet e as extracted from IFS data. IRDIS photometric points are overplotted. A comparison with an L6 isolated brown dwarf shows a mismatch at $K$ band wavelengths. { There is a slight mismatch between IRDIS and IFS measurements, as we separately normalized  the flux of the two detectors with the corresponding PSF, the offset could be induced by another variable. }}
\label{f:sp_e}
\end{center}
\end{figure*}

\begin{figure*}
\begin{center}
\includegraphics[width=1.3\columnwidth]{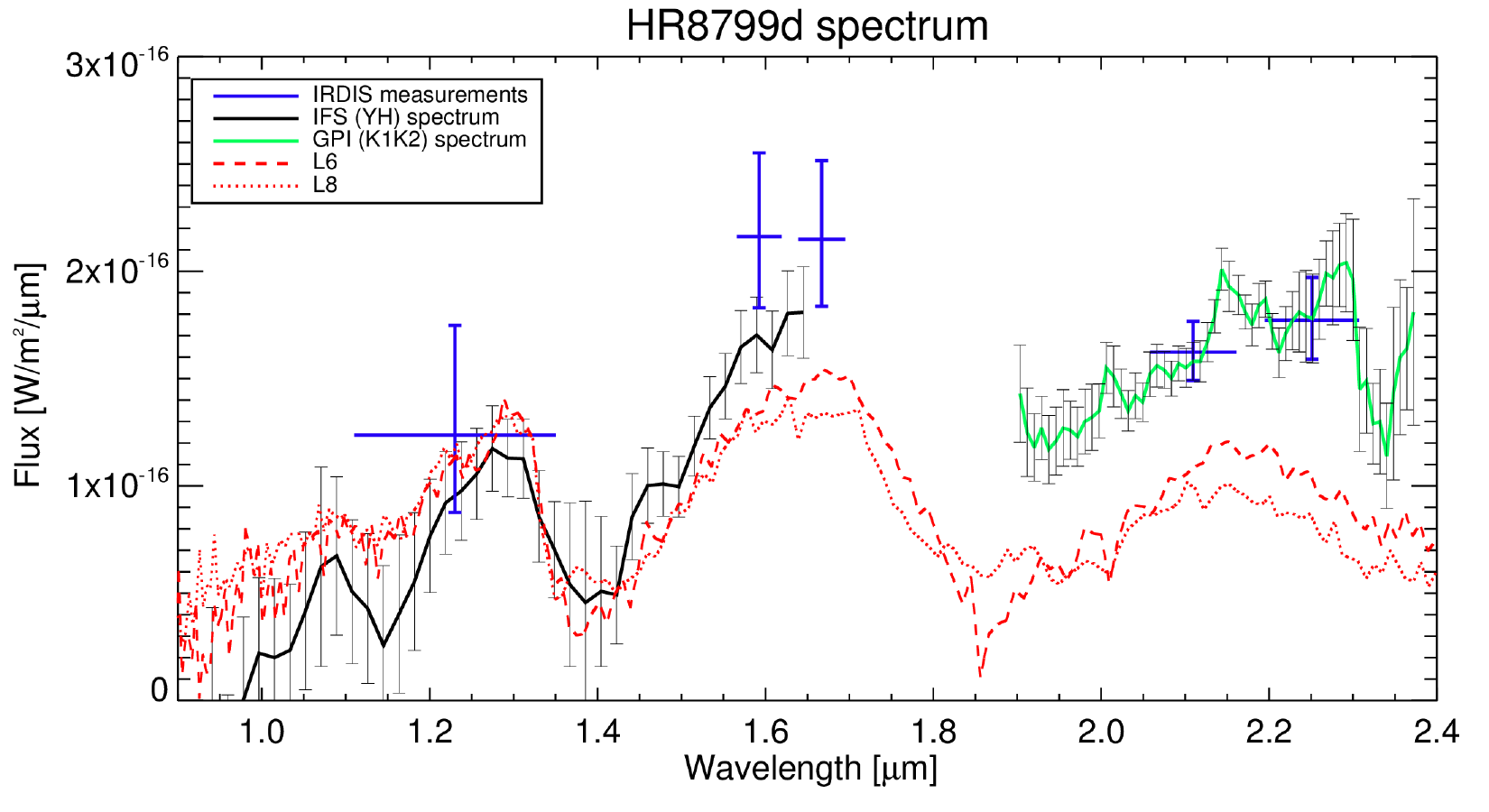}
\caption{Spectrum of planet d as extracted from IFS data. IRDIS photometric points are overplotted. The GPI spectrum is also included in the plot.  Comparison with a L6-L8 isolated brown dwarf shows a mismatch at $K$-band wavelengths. { There is a slight mismatch between IRDIS and IFS measurements, as we normalized separately the flux of the two detectors with the corresponding PSF, the offset could be induced by another variable.}}
\label{f:sp_d}
\end{center}
\end{figure*}

  We used this flux-calibrated model spectrum of HR\,8799\,A to retrieve the flux  of  HR\,8799\,bcde in the IRDIS passbands (Table~\ref{TabApA:HR8799b}) and to obtain flux-calibrated spectra of the two innermost planets.   Figures~\ref{f:sp_e} and \ref{f:sp_d} show that the $J$, $H2$, and $H3$ photometry of HR\,8799\,d and e is consistent with the spectra extracted from the IFS datacubes.  We also included the flux-calibrated $K$-band GPI spectrum of HR\,8799\,d obtained by \cite{2014ApJ...794L..15I}. This spectrum is compatible with IRDIS $K1$ and $K2$ photometry of the planet.

\begin{figure*}
\begin{center}
\includegraphics[width=1.3\columnwidth]{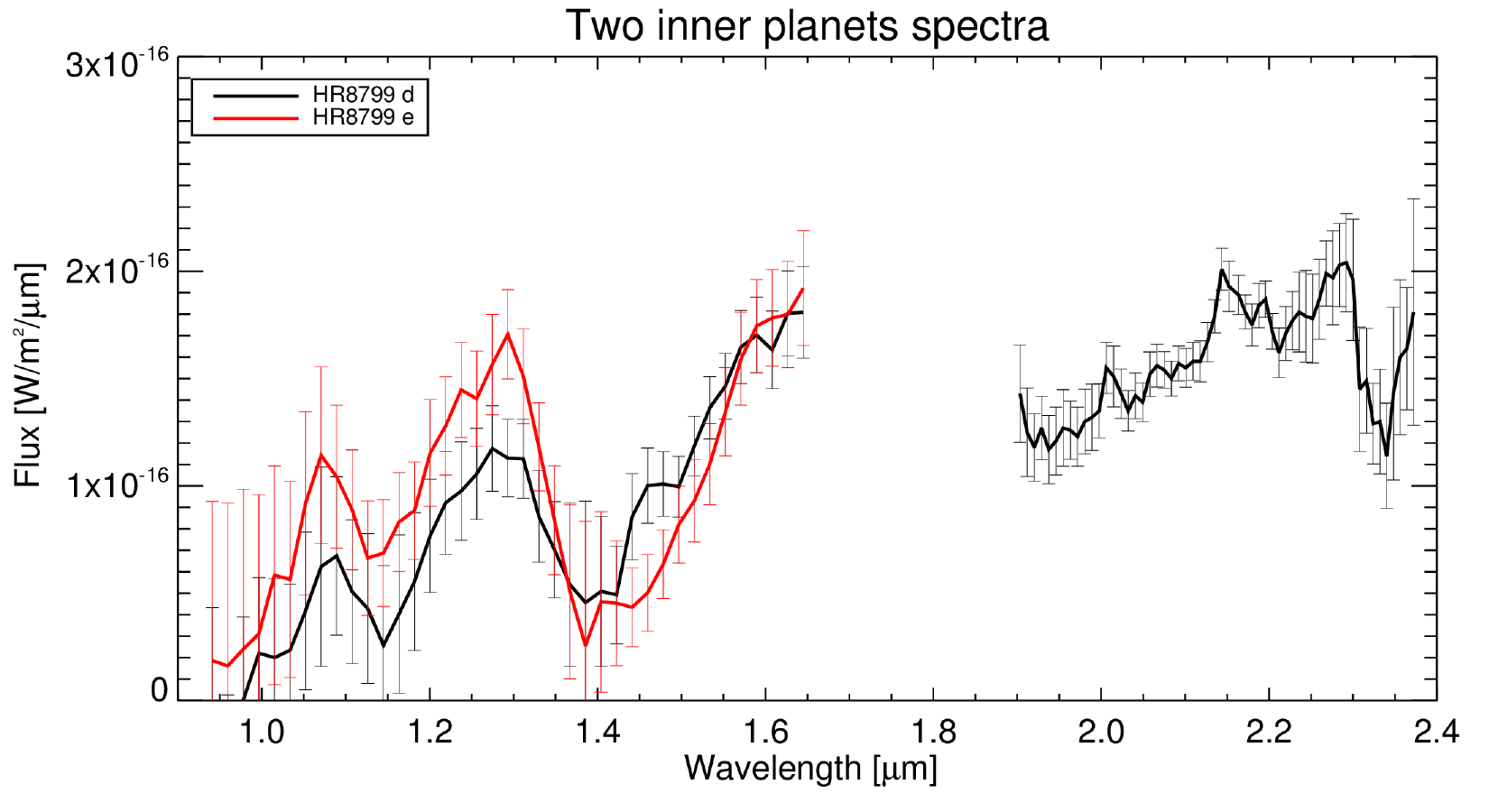}
\caption{Spectra of the two planets HR\,8799\,de together.}
\label{f:twopl}
\end{center}
\end{figure*}

  { The spectra of the two planets together are shown in Fig.~\ref{f:twopl}.}

 \begin{table*}[t]
\caption{Currently available fluxes of HR\,8799\,b at 10~pc gathered from narrowband and broadband photometry.}
\label{TabApA:HR8799b}
\begin{center}
\renewcommand{\footnoterule}{}  
\begin{tabular}{cccccc}
\hline \hline 
Filter &$\lambda$ & FWHM & Abs. Flux & err. Flux + & err. Flux - \\
 &($\mu$m) &($\mu$m) &(Wm$^{-2}\mu$m$^{-1}$) & (Wm$^{-2}\mu$m$^{-1}$) &(Wm$^{-2}\mu$m$^{-1}$) \\ 
\hline                          
\hline
\multicolumn{6}{c}{HR\,8799\,b}\\
\hline
\hline

BB\_$J$ & 1.25  &0.20   &6.42E-16       &6.72E-17       &6.09E-17       \\
 $H2$ &1.59     &0.05   &1.26E-15       &1.90E-16       &1.65E-16       \\
 
 $H3$ &1.67     &0.06   &1.42E-15       &1.61E-16       &1.44E-16       \\

 $K1$ &2.11     &0.10   &1.05E-15       &8.54E-17       &7.90E-17       \\
 $K2$ &2.25     &0.11   &9.71E-16       &1.02E-16       &9.21E-17       \\
\hline
\hline
\multicolumn{6}{c}{HR\,8799\,c}\\
\hline
\hline                          

BB\_$J$         &1.25&  0.20&   1.90E-15&       2.68E-16&       2.35E-16        \\
$H2$            &1.59&  0.05&   3.15E-15&       4.14E-16&       3.66E-16        \\
$H3$            &1.67&  0.06&   3.56E-15&       4.03E-16&       3.62E-16        \\
$K1$    &       2.11&   0.10&   2.54E-15&       1.90E-16&       1.76E-16        \\
$K2$    &       2.25&   0.11&   2.70E-15&       2.39E-16&       2.20E-16        \\
\hline
\hline
\multicolumn{6}{c}{HR\,8799\,d}\\
\hline
\hline                          

BB\_$J$         &       1.25&   0.20&    1.92E-15&      7.92E-16&       5.61E-16        \\
$H2$            &       1.59&   0.05&    3.35E-15&      6.06E-16&       5.13E-16        \\

$H3$            &       1.67&   0.06&    3.34E-15&      5.69E-16&       4.86E-16        \\

$K1$            &       2.11&   0.10&    2.52E-15&      2.23E-16&       2.05E-16        \\

$K2$            &       2.25&   0.11&    2.75E-15&      3.12E-16&       2.80E-16        \\

\hline
\hline
\multicolumn{6}{c}{HR\,8799\,e}\\
\hline
\hline                          
BB\_$J$         &1.25&  0.20&   2.29E-15&       5.10E-16&       4.17E-16        \\
$H2$    &       1.59&   0.05&   3.71E-15&       7.87E-16&       6.50E-16        \\
$H3$    &       1.67&   0.06&   3.90E-15&       8.69E-16&       7.11E-16        \\

$K1$    &       2.11&   0.10&   2.71E-15&       3.07E-16&       2.76E-16        \\
$K2$    &       2.25&   0.11&   2.80E-15&       3.42E-16&       3.05E-16        \\

\hline
\end{tabular}
\end{center}
 \end{table*}

\begin{figure}
\begin{center}
\includegraphics[width=\columnwidth]{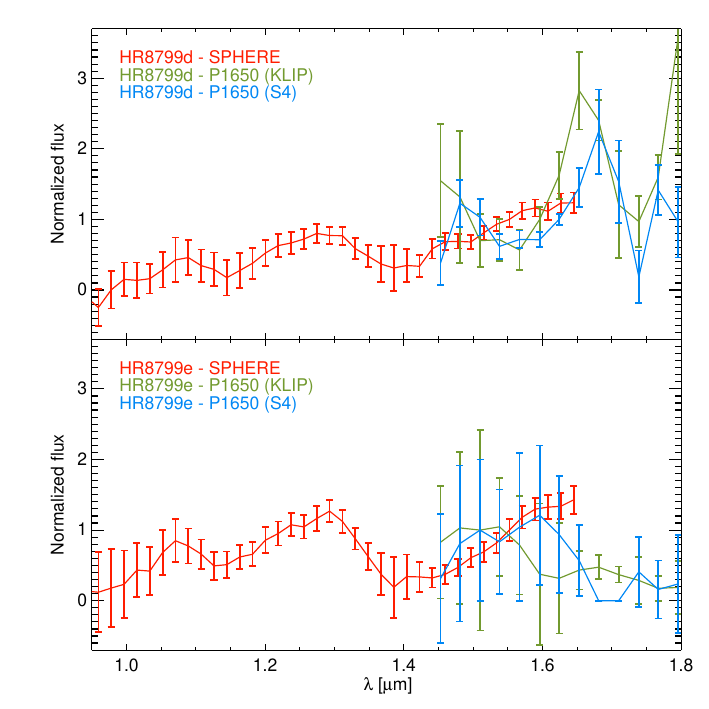}
\caption{Comparison of HR\,8799\,de spectra retrieved from SPHERE data and P1640 data \citep{2013ApJ...768...24O}.}
\label{f:comp}
\end{center}
\end{figure}

A comparison of the two spectra with results from Project 1640 \citep{2013ApJ...768...24O} is shown in Fig.~\ref{f:comp}. A table with the flux values for each IFS channel is available in the electronic version of this paper.

We compared the flux-calibrated spectra, $H2$, $H3$, $K1$, and $K2$ fluxes of HR\,8799\,d and e to those of field brown dwarfs from \citet{Testi2001} and to 1000 empirical templates from the SpeXPrism library. While the \cite{Testi2001} library mostly contains spectra of old objects, the SpeXPrism library includes spectra of some { dusty} dwarf and peculiar (red) L and T dwarfs. The comparison { illustrates, as previously found in the literature,} that the planet properties are best reproduced by objects at the L/T transition. The spectral shapes of the SPHERE/IFS spectra of the two innermost planets are reproduced by those of L6 field dwarfs (Fig.~\ref{f:sp_e} and \ref{f:sp_d}). Nevertheless, the planet spectra have higher fluxes than these field dwarfs at longer wavelengths (i.e. their spectral slopes are redder).

Among the sample of  SpeXPrism spectra,  those of  \object{2MASS J2244\-3167+2043\-433} \citep[L7.5/L6.5,][]{2002AJ....124.1170D}  and \object{SDSS J151643.01+305344.4} \citep[T0.5-1.5pec,][]{2006AJ....131.2722C, 2010ApJ...710.1142B} best match the spectra of HR8799d and e, respectively. \cite{2011ApJ...733...65B} has already noted the similarity of SDSS 1516+30 to HR\,8799\,b. Both SDSS 1516+30 and 2MASS 2244+20  are shown in Figures \ref{f:JH2K2K1} and \ref{f:JK1H2K1}. The colors of \object{2MASS J2244\-3167+2043\-433,} along with the peculiar L6 dwarf \object{2MASS J21481633+4003594} \citep[][]{2008ApJ...686..528L}, deviate from the sequence of field dwarfs. These objects are red dusty dwarfs proposed to be young members of the field population \citep[][and ref. therein]{2009ApJ...702..154S, 2014ApJ...783..121G}.

In summary, the analysis { demonstrates that there is a} deviation of HR\,8799\,bcde spectrophotometric properties from the sequence of field dwarfs. The properties of HR\,8799\,d and e are shared by some red dusty, and possibly young, objects at the L/T transition. We investigate these properties in more detail in the companion paper, Bonnefoy et al., { A\&A accepted}.

 \section{Conclusions}
\label{sec:conc}

We present first results of the planet hunter
SPHERE on the planetary mass objects surrounding the young intermediate mass star HR\,8799. We observed this system during the commissioning and the science verification runs
of SPHERE during the months of July, August, and December
2014. Four previously known planets have been detected in $J$, $H2H3$, $H$, and
$K1K2$ bands with high SNR (reaching $\sim$ 200 in $K$ band). For the
first time, planet HR\,8799\,e has been detected in $J$ band.
Optimized post-processing methods were exploited in the analysis of these data.
 We used 
independent pipelines based on KLIP and T-LOCI, with the injection of fake negative companions to have robust measurements of photometry, spectrometry, and astrometry.
We obtained photometry in the four dual- and broadband filters as well as spectra in $YH$ band for HR\,8799\,de. 
The quality of these spectra, with mean SNR $\sim$ 20,
make them the highest quality spectra yet obtained for these objects.

The astrometric positions of the four planets are given for the
epochs of July and December. For the closest planets d and e, we
also provide astrometric positions of August 2014, which were
extracted from IFS data. For this epoch, given the high SNR ratio
and the small pixel scale of IFS, we are able to obtain astrometric
positions with error bars of { 2-4 mas}.

{ We find that two orbital solutions are the most consistent with our measurements for the two inner planets, corresponding to the resonance 2d:1e and 3d:2e.}

{ We found that a possible inner planet f, which is expected to lie on the 3e:1f and 2e:1f resonance orbits, would be detected by IFS with masses greater than 3-7 \MJup.} { Following previous stability simulations, we estimated that for the 3e:1f orbit 2-4.5 \MJup planets are not excluded, and in the 2e:1f orbit we do not exclude the presence of 1.5-3 \MJup planets.}

Finally, we demonstrate that the spectrophotometric properties of HR\,8799\,d and e in the $YH$ range are similar to those
of brown dwarfs with spectral type L6-L8. These planets show redder colors than field objects in the L-to-T transition.  Empirical spectra of old brown dwarfs cannot describe the spectral energy distribution of these objects at longer wavelengths, { and to comprehend the mechanisms that cause this red excess, further analysis is needed.}

\begin{acknowledgements}
 We acknowledge the two anonymous referees for their constructive comments. We are grateful to the SPHERE team and all the people at Paranal for the great effort during SPHERE commissioning runs. We especially thank A. Bellini and J. Anderson for the HST astrometric measurements. { A.Z. acknowledges support from the Millennium Science Initiative (Chilean Ministry of Economy),
through grant ``Nucleus RC130007''}. D.M., A.-L.M., R.G., R.U.C., S.D. acknowledge support from the ``Progetti Premiali'' funding scheme of the Italian Ministry of Education, University, and Research. We acknowledge support from the 
French National Research Agency (ANR) through the GUEPARD project grant ANR10-BLANC0504-01 and the Programmes Nationaux de Planetologie et de Physique Stellaire (PNP \& PNPS) . Part of this work has been carried out within the frame of the National Centre for Competence in
Research PlanetS supported by the Swiss National Science Foundation. S.P.Q. and M.R.M. acknowledge the
financial support of the SNSF. The author thanks Sebasti\'an Per\'ez for his help in solving unsolvable latex problems.
\end{acknowledgements}

\begin{tiny}
SPHERE is an instrument designed and built by a consortium consisting
of IPAG (Grenoble, France), MPIA (Heidelberg, Germany), LAM (Marseille,
France), LESIA (Paris, France), Laboratoire Lagrange (Nice, France), INAF
- Osservatorio di Padova (Italy), Observatoire de Genève (Switzerland), ETH
Zurich (Switzerland), NOVA (Netherlands), ONERA (France), and ASTRON
(Netherlands), in collaboration with ESO. SPHERE was funded by ESO, with
additional contributions from CNRS (France), MPIA (Germany), INAF (Italy),
FINES (Switzerland), and NOVA (Netherlands). SPHERE also received funding
from the European Commission Sixth and Seventh Framework Programmes as
part of the Optical Infrared Coordination Network for Astronomy (OPTICON)
under grant number RII3-Ct-2004-001566 for FP6 (2004-2008), grant number
226604 for FP7 (2009-2012) and grant number 312430 for FP7 (2013-2016).
\end{tiny}

 
\bibliographystyle{aa}
\bibliography{hr8799}

\end{document}